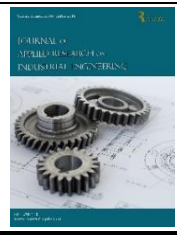

Paper Type: Original Article

# Leveraging Deep Learning Techniques on Collaborative Filtering Recommender Systems

**Ali Fallahi RahmatAbadi** [1], **Javad Mohammadzadeh** [2] *

[1] Department of Computer Engineering, Karaj Branch, Islamic Azad University, Karaj, Iran; ali.fallahi@kiau.ac.ir;
[2] Department of Computer Engineering, Karaj Branch, Islamic Azad University, Karaj, Iran; j.mohammadzadeh@kiau.ac.ir;

## Abstract

With the exponentially increasing volume of online data, searching and finding required information have become an extensive and time-consuming task. Recommender Systems as a subclass of information retrieval and decision support systems by providing personalized suggestions helping users access what they need more efficiently. Among the different techniques for building a recommender system, Collaborative Filtering (CF) is the most popular and widespread approach. However, cold start and data sparsity are the fundamental challenges ahead of implementing an effective CF-based recommender. Recent successful developments in enhancing and implementing deep learning architectures motivated many studies to propose deep learning-based solutions for solving the recommenders' weak points. In this research, unlike the past similar works about using deep learning architectures in recommender systems that covered different techniques generally, we specifically provide a comprehensive review of deep learning-based collaborative filtering recommender systems. This in-depth filtering gives a clear overview of the level of popularity, gaps, and ignored areas on leveraging deep learning techniques to build CF-based systems as the most influential recommenders.

**Keywords**: Recommendation Systems, Deep Learning Architectures, Collaborative Filtering, Survey

## 1 | Introduction

As a consequence of the exponential growth of online-based technologies, we have experienced many significant changes in various aspects of our life. In the digital age, individuals can store and access information in ways that were not accessible in the past. However, these brilliant opportunities caused some critical challenges. Searching for a specific record inside the continuously increasing amount of data becomes more complicated and confusing. As a subclass of decision support systems, recommender systems help users overcome the information overload and access what they need rapidly through personalized channels [1].

In the last decades, working with big data was a challengeable concern. However, today, extensive datasets are valuable resources for complex systems that work based on deep structured learning methods because of the recent development in artificial intelligence and computation power. Strength capabilities of deep learning-based approaches compared with traditional techniques for solving complicated problems caused significant attention to employing deep learning in various domains such as image processing [2], speech recognition [3], data mining [4], business [5], natural language processing [6], and information filtering systems like recommendation engines [7].

The recommender system's primary intention is to predict the user's tendency toward an item; the item can be a company's product, stock in a stock market, a friend in a social network [8], a movie, or a photo on a website, etc [9]. Based on this concept, many researchers proposed various recommenders with diverse functionalities to suggest books, films, music, hotels, friendships, ps, etc. [10; 11].

## 1.1 | Traditional Approaches to Build a Recommender System

Recommender systems can mainly be categorized into three types: Content-based, Collaborative filtering, and Hybrid recommenders [12]. We illustrated this categorization in *Fig. 1*, which shows an overview of the three mentioned criteria. In the next following paragraphs, we will explain and pinpoint some notable aspects of each type of recommender system.

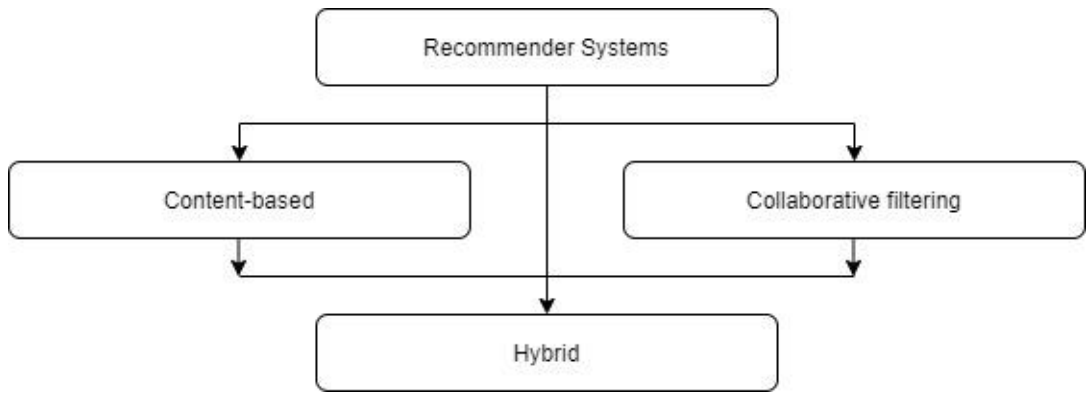

**Fig. 1. Recommender systems categorization based on their techniques.**

### 1.1.1 | Content-based

In this technique, the recommender engine considers the features of what users chose in the past to make suggestions for other related items of the dataset. Textual descriptions and tags are typical resources for implementing a content-based recommender [13; 14]. For instance, in a content-based book recommender system, if a user likes a book about deep learning, the system will first analyze the book's available textual properties, such as title, author, genres, complete text, etc. Then, various techniques like keyword-based vector-space structure will be used to find the other most similar books in the dataset with the highest similarity matching score with the chosen deep-learning book [15].

### 1.1.2 | Collaborative filtering

The fundamental hypothesis of Collaborative Filtering is that users who had similar tastes and behaviors in the past will also have similar activities in the future. This strategy uses ratings or other measurable user's activities such as positive/negative comments, like/dislike, etc., to find similar users and then provide recommendations based on their similarities. Among all of the mentioned techniques, collaborative filtering is the most well-known and popular method for implementing a recommender system [16; 17]. We describe the Collaborative Filtering technique in detail in the following sections. There are two main approaches to build a Collaborative Filtering system which are memory-based and model-based. The memory-based approach utilizes user rates to calculate the correlation among users or items. In a model-based approach, the primary step is to use the dataset to learn a proposed model. In the next step, the model is applied for making predictions. Matrix Factorization is the most common algorithm in building model-based Collaborative Filtering systems [18].

### 1.1.3 | Hybrid recommenders

The critical factor in building an efficient recommender system is to improve accuracy and provide more personalized suggestions. Some studies tried to create hybrid systems based on a mixture of other techniques to benefit from the basic methods such as collaborative filtering and content-based and overcome their drawbacks [19; 20]. Recent achievements in deep learning and neural networks also provided new opportunities for building hybrid systems that handle large amounts of data on complex networks [21; 22]. Hybridization can be done in different types. For instance, the system can provide recommendations based on various features generated by basic recommenders; this method is known as feature combination. Another approach is Switching; the recommender system switches within different ways to provide recommendations [23].

## 1.2 | Related Studies

In the past few years, some studies such as Khan et al. [7], Da'u and Salim [9], Zhang et al [10], Batmaz et al. [16] were conducted to review and survey deep learning-based recommender systems. However, to the best of our knowledge, not a single study specifically focused on leveraging deep learning techniques on collaborative filtering recommenders as the most common technique to build recommender systems [24]. In the following lines, we introduce mentioned researches more by explaining their notable aspects.

Khan et al. [7] provided a comprehensive survey about deep learning-based rating prediction approaches. The authors reviewed different algorithms and architectures by concentrating on rating prediction systems; however, the main difference between the study and our research is providing an in-depth review by focusing on collaborative filtering recommenders built based on deep learning architectures. By emphasizing presenting a systematic literature review (SLR), Da'u and Salim [9] provided a survey about building a recommender system based on deep learning techniques. Zhang et al [10] mainly focused on the taxonomical classification of reviewed studies and their approaches. Batmaz et al. [16], to help future researchers interested in the topic, categorized reviewed publications based on four dimensions: deep learning models and architecture, possible solutions for the challenges, recommender application domains, and purposive properties.

To provide a thorough study, we also reviewed acclaimed surveys about deep learning architectures, which were not limited to the subject of recommendation systems. For instance, Shrestha and Mahmood [25] proposed a review of deep learning algorithms and architectures by focusing on mathematical concepts of enhancing training operations. Although the study is not written on recommender systems, the authors flawlessly explained details about the structure of different deep learning architectures.

As a contribution, unlike past researches about using deep learning architectures in recommender systems that covered different techniques generally, in this study, we specifically provide a comprehensive review of deep learning-based collaborative filtering recommender systems to guide and assist new researchers interested in the area.

The rest of the paper is organized as follows: Section 2 provides the preliminaries of recommender systems include traditional approaches and fundamental challenges. Deep learning-based recommender systems are discussed in Section 3. Section 4 discusses the details of our results from some of the essential views applied to the topic. In Section 5, we present our conclusions and future work.

## 2 | Background

This section describes the collaborative filtering technique with more details as the main focus of this study and the most commonly utilized method to build recommender systems. Moreover, in the following, we explain fundamental challenges ahead of implementing an efficient, accurate recommender.

### 2.1 | Collaborative Filtering Recommenders in Detail

In comparison with the content-based approaches, in collaborative filtering, the system can provide recommendations based on the similarity of user's activities (or items' characteristics) without the necessity of analyzing the items [26]. *Fig. 2*, shows a user-item rating matrix. This is a sample scenario of how a collaborative filtering recommender engine predicts a specific user's rates for an item. In this example, ratings are between one to five. We selected Alice as the target user. The system aims to predict Alice's possible ratings for items that she did not rate. Items can be imagined as movies that she did not watch. Then recommend the items with the highest rates to her.

|       | Item1 | Item2 | Item3 | Item4 | Item5 | Item6 | Item7 |
|-------|-------|-------|-------|-------|-------|-------|-------|
| Alice | 4     |       |       | 5     | 1     |       |       |
| Bob   | 5     | 5     | 4     |       |       |       |       |
| Jim   |       |       |       | 2     | 4     | 5     |       |
| Kate  |       | 3     |       |       |       |       | 3     |

Fig. 2. A view of the User-Item matrix.

The following paragraphs clarify the method step by step until proving the recommendation. The first step is calculating the similarity between Alice and the other three users. There are different metrics [27] for calculating similarity in recommender systems, such as:

#### 2.1.1 | Jaccard similarity

In the below formula, the Jaccard Similarity of users $p$ and $q$ is calculated from the number of items rated both by the users $p$ and $q$ divided by the union of rated items by the users $p$ and $q$. The fraction's numerator can be defined as the total number of the co-rated items between the users $p$ and $q$. The denominator also can be defined as the total number of the rated items by both the users, $p$ and $q$ [28].

$$\text{JaccrdSim}_{p,q} = \frac{\left| r_p \cap r_q \right|}{\left| r_p \cup r_q \right|} \qquad (1)$$

*Table 1* shows the result of calculating the Jaccard similarity between Alice and other users in the dataset.

Table 1. Result of the Jaccard similarity between Alice and other users.

|       | Bob | Jim | Kate |
|-------|-----|-----|------|
| **Alice** | 1/5 | 2/4 | 0    |

The above results indicate that Alice has the most similarity with Jim. However, the Jaccard similarity's main drawback is that the metric ignores how much two users are similar and only counts the number of co-rated items, whether the ratings are identical or the opposite [29].

### 2.1.2 | Cosine similarity

Cosine similarity measures similarity by calculating the cosine angle between the two rating vectors given by two targeted users. The smaller value of angle represents higher similarity and vice versa [30]. Cosine similarity is calculated as follows:

$$\text{Cos}_{p,q} = \frac{\overline{R_p} \cdot \overline{R_q}}{\left|\overline{R_p}\right| \cdot \left|\overline{R_q}\right|} \quad (2)$$

In the Cosine similarity formula *Eq. (2)*, *Rp* and *Rq* respectively represent rating vectors of user *p* and *q*. In the numerator of the fraction, ".", indicates the dot product of two vectors.

To employ the Cosine similarity for the mentioned example, there should be some values for unrated items. The simplest way to complete this step is by adding zero to the empty cells. *Fig. 3*, shows the user-item matrix after adding zero values to the empty cells to calculate the Cosine similarity between Alice and other users.

|       | Item1 | Item2 | Item3 | Item4 | Item5 | Item6 | Item7 |
|-------|-------|-------|-------|-------|-------|-------|-------|
| Alice | 4     | 0     | 0     | 5     | 1     | 0     | 0     |
| Bob   | 5     | 5     | 4     | 0     | 0     | 0     | 0     |
| Jim   | 0     | 0     | 0     | 2     | 4     | 5     | 0     |
| Kate  | 0     | 3     | 0     | 0     | 0     | 0     | 3     |

Fig. 3. User-Item matrix after adding zero values.

*Table 2* shows the result of calculating the Cosine similarity between Alice and other users in the dataset

Table 2. Result of the Cosine similarity between Alice and other users.

|  | Bob | Jim | Kate |
|---|---|---|---|
| **Alice** | 0.93 | 0.75 | 0 |

In contrast to the Jaccard similarity, the results of the Cosine similarity indicate that Alice decides more similarly to Bob than Jim. By looking at the rating scores, it can be concluded Cosine similarity results are more realistic. However, treating missing ratings the same as the negative rates is a disadvantage of the Cosine similarity. In the example, we set all the empty cells with zero; in other words, we assigned an uncertain rate for unrated items, which can be utterly wrong. To clarify the problem, if, in a movie recommender, the system sets zero for unrated movies, the user may give it a high rating after watching it.

A solution to overcome this problem is to use the Centered Cosine. The Centered Cosine's main idea is to normalize ratings by subtracting each rate from the average of those ratings for the target user. Based on the explained situation, the concept in Centered Cosine similarity can be considered similar to the Pearson Correlation Coefficient.

### 2.1.3 | Pearson correlation coefficient

Pearson Correlation Coefficient (PCC) [31] is one of the most widespread and notable popular similarity measure recommenders [32]. *Eq. (3)* shows the PCC formula.

$$\text{PCC}_{a,b} = \frac{\sum_{i=1}^{I_{a,b}} (r_{a,i} - \bar{r}_a)(r_{b,i} - \bar{r}_b)}{\sqrt{\sum_{i=1}^{I_a} (r_{a,i} - \bar{r}_a)^2} \sqrt{\sum_{i=1}^{I_b} (r_{b,i} - \bar{r}_b)^2}} \quad (3)$$

In the PCC formula, *Eq. (3)*, $r_{a,i}$ indicates the rating score for item i, from the target user a. $r_{b,i}$ denotes the rating score for the same item from the user b. $\bar{r}_a$ and $\bar{r}_b$ mean the average rating of user *a*, and user *b* based on all rated items by each user.

|  | Item1 | Item2 | Item3 | Item4 | Item5 | Item6 | Item7 |
|---|---|---|---|---|---|---|---|
| Alice | 2/3 |  |  | 5/3 | -7/3 |  |  |
| Bob | 1/3 | 1/3 | -2/3 |  |  |  |  |
| Jim |  |  |  | -5/3 | 1/3 | 4/3 |  |
| Kate |  | 0 |  |  |  |  | 0 |

**Fig. 4. The modified User-Item matrix after subtracting the rates of each row from its average.**

*Fig. 4*, shows the modified version of the user-item matrix after subtracting each rate from the average of ratings in that row. The final result of *Eq. (3)* is shown in *Table 3*.

Table 3. Result of the Pearson Correlation Coefficient similarity between Alice and other users.

|  | Bob | Jim | Kate |
|---|---|---|---|
| **Alice** | 0.97 | - 0.57 | 0 |

By using the presented results in *Table 3*, the difference between Alice and Jim is more straightforward. So, Bob is the most similar user to Alice, and as can be seen, PCC captures the intuition better.

Typically, the second step of the collaborative filtering is selecting Top-N most similar users to the target user. The concept is known as the k-nearest neighbor method, and the main idea is to categorize other users' similarity values based on a predefined threshold [33]. Neighbors who, their score is greater than the threshold will be assigned to the Top-N group [34; 35]. The variable *N* can be set with different values. It had to be mention that all the members of this group must have rated the target item. In the above example, based on the PCC's result, it can be concluded that Bob is the most similar user to Alice in the dataset.

The final step is predicting the target user's rate for the target item. There are different formulas to do this step. However, to complete the example, we chose the average rating prediction model shown in *Eq. (4)* to provide predictions.

$$r_{x,i} = \frac{1}{k \sum_{y \in N^r Y^i}} \qquad (4)$$

In the equation *Eq. (4)*, $r_{x,i}$ is the predicted rate for item *i* from user *x*, which calculates the average rating of all the users in the selected neighborhood for item *i*. Also, the set *N* consists of users who rated item *i* and are similar to user *x*. Obviously, based on the total number of users in the presented example, Bob is the only similar user to Alice after selecting the most Top-N similar users. So, the predicted scores for unrated items by Alice will be similar to Bob's rates.

## 2.2 | Fundamental Challenges

Recommender systems are faced with different challenges. The increasing population of online users and numerous items caused difficulties such as sparsity, cold start, and the Grey sheep problem [36; 37]. This section outlines the mentioned challenges in making an efficient and accurate recommender system.

### 2.2.1 | Sparsity

Among the mentioned techniques for building a recommender, the collaborative filtering method has a high dependency on the user's interaction with the system. However, in most datasets, there is a lack of sufficient data about users and items such as rates, comments, reviews, likes, dislikes, etc. Solving the sparsity problem was an appealing subject for many studies [38; 39]. While explicit trust relationships are used in many studies as a reliable approach to alleviating the data sparsity problem, some studies introduced propagation of trust and distrust values as a more practical solution to explore the unstated relationships between users. As a result of these activities, a collaborative filtering recommender has more data to calculate similarities and provide suggestions [40].

### 2.2.2 | Cold start

Cold start happens when a new user or a new item recently joined a system. There is not enough information about the user's activities in the past or enough interactions and feedback about the new item in this situation. The cold start has an extensive negative effect in collaborative filtering-based recommender systems, since the recommender engine does not have enough information or feedback to calculate similarity [41]. Considering other origins of information such as social networks or contextual and demographic data is an efficient solution to overcome the cold start problem. In this regard, Linked Open Data and DBpedia are two valuable resources to gain more information about users or items [42].

### 2.2.3 | Grey sheep

The problem of Grey Sheep users is a severe challenge of collaborative filtering recommender systems. The "gray sheep" term refers to the users who are not similar to the majority of other users. This issue makes it difficult for recommender engines, especially the collaborative filtering ones, to calculate the similarity between users and provide accurate suggestions [43; 44]. Researchers also tried to propose modern solutions to overcome the Grey Sheep problem with the development of machine learning techniques. For instance, using clustering algorithms is an effective solution to identify Grey Sheeps. Another approach is extracting content-based features from the Grey Sheep user's profile to improve recommendations' accuracy [45].

## 3 | Deep learning-based Recommender Systems

In recent years studying the influence of deep learning in different areas attained substantial interest. Likewise, in recommender systems, employing deep-learning techniques helped the experts enhance previous achievements and provide more accurate and precise results by prevailing over the fundamental challenges such as data sparsity, cold start, and grey sheep users [45; 46].

### 3.1 | Main Deep Learning Architectures in Collaborative Filtering Recommender Systems

In contrast to the past studies about applying deep learning architectures in recommender systems that made a general overview of different deep learning approaches, in this section, we expressly present a comprehensive analysis of deep learning-based collaborative filtering recommender systems.

### 3.1.1 | Restricted boltzmann machines

Restricted Boltzmann machine (RBM) is a special kind of Boltzmann machine. The RBM makes it possible to detect patterns in the input data by reconstructing them automatically. RBM is a network was built from two layers. Layers, respectively, are named visible and hidden layers. Each node in the first layer has a link with all the nodes in the hidden layer. The model is considered restricted because there is no connection between the nodes in the same layer [47]. *Fig. 5*, shows an illustration of the RBM architecture.

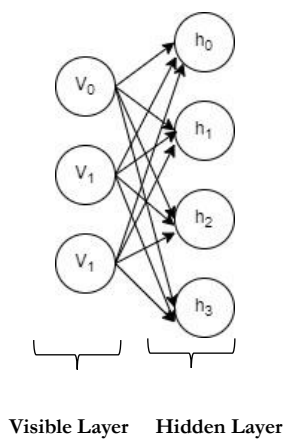

**Fig. 5. A Restricted Boltzmann Machine architecture.**

Louppe [48] used parallel computing techniques with shared memory, distributed computing, and method ensembles for providing RBM-based collaborative filtering systems. The author's experimental results indicate that parallel computing can be an effective solution to improve the provided suggestion's accuracy.

Georgiev and Nakov [49] introduced a joined user-based and item-based collaborative filtering in a unified framework based on RBM. Moreover, the researchers employed real data in the first layer of the RBM architecture instead of multinominal variables. The authors also investigated the probability of mixing the RBM-based method's knowledge and the actual information.

Liu et al. [50] proposed a hybrid model based on RBM architecture and a collaborative filtering approach. The authors used items' categories as the system's input to enhance the system performance and increase the result's accuracy.

Zheng et al. [51], introduced a collaborative filtering Neural Autoregressive Distribution Estimation model named CF-NADE, which provides recommendations using RBM architecture. Authors showed that leveraging a deep learning network such as RBM can enhance the traditional and basic collaborative filtering approaches.

Jia et al. [52] proposed a collaborative-based RBM recommender system for exhibition managements and participators in social events. The introduced recommendation framework mixes the data from various references and builds a relationship among the online knowledge and users who participated in the target event.

Du et al. [53] introduced an item-based RBM method for collaborative filtering and applied the deep multilayer RBM network structure to overcome the sparsity problem. The authors considered every item as a separated RBM while each machine has similar properties such as weights and biases. The parameters are learned layer by layer in the deep network. They also used the batch gradient descent algorithm with minibatch to boost the convergence speed.

Wu et al. [54], to enhance the recommendations, considered trust relationships in recommender systems. The authors utilized explicit trust values and user's ratings as input data of the machine and proposed a social recommendation technique based on RBM.

### 3.1.2 | Autoencoders

Autoencoders are a neural network that takes an unlabeled set of inputs and reconstructs accurate results after encoding them. The system acts as a feature extraction engine and decides which data features are the most important. Autoencoders are generally a shallow network and consist of three layers: input, hidden, and output layers. A RBM can be considered as a two-layers Autoencoder. The system has two general steps, which are encoding and decoding. Typically, the features used to encode input for the hidden layer, also used for decoding and provide results in the output layer. The process of the forward propagation from the input layer toward the output layer and the backward propagation in a reverse path is repeated continuously to achieve acceptable accuracy. *Fig. 6*, shows an illustration of the RBM architecture [55].

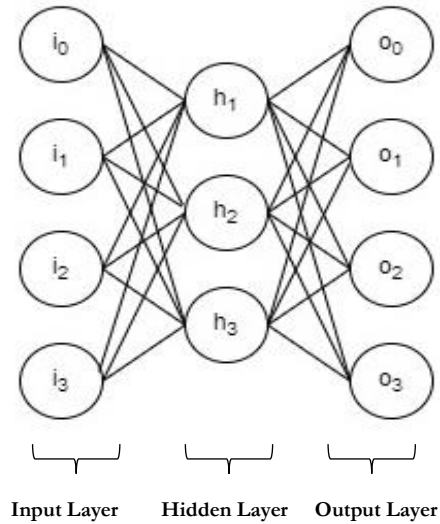

**Fig. 6. An Autoencoder architecture.**

Ouyang et al. [56] introduced one of the most pioneer studies in the autoencoder-based collaborative recommendation model. The authors utilized RBM and autoencoders together in the pretraining phase to achieve more personalized and accurate results. The system utilized a non-linear form of the input data.

Li et al. [57] proposed a recommender system, based on the mix of probability Matrix Factorization in collaborative filtering with Autoencoder architecture. Due to their results, the model has significant efficiency in working with massive datasets such as movie or book recommenders.

Sedhain et al. [58] introduced the AutoRec recommender that uses user and item vectors based on Autoencoders architecture and collaborative filtering approach.

Strub and Mary [59] proposed an autoencoders structure that calculates a non-linear matrix factorization based on the sparse ratings as inputs of the system.

Wang et al. [60] employed a stacked denoising autoencoders architecture to enhance rating prediction accuracy. The introduced model leverage content data and a collaborative filtering approach based on user-item rating matrix values.

Wang et al. [61] implemented a collaborative recurrent autoencoder (CRAE) that considers textual information and rating values. The authors technically used the hierarchical Bayesian model for denoising recurrent autoencoder.

Ying et al. [62] proposed a mixed model of a pair-wise recommender system that considers unexplicit feedback and collaborative filtering concept. The authors employed Stacked Denoising Autoencoders (SDAE) to select the item description's characteristic features. The system utilized a Bayesian framework to combine rates and other data about items.

Wei et al. [63] proposed a recommender system based on collaborative filtering and SDAE architecture to overcome the cold-start problem. The inputs of the model are the item's textual properties and user choices, and activities. An extended version of this research is proposed in [64].

Suzuki and Ozaki [65] employed users' ratings as inputs for an autoencoder architecture and computed the similarity between users in hidden layers. The decoded output of the system is a predicted rating used to provide recommendations.

Li and She [66] proposed a Bayesian generative multimedia recommender system based on a collaborative variational autoencoder. The system indicated both rating and content to explore the implicit connection between users and items.

Liang et al. [67] proposed a collaborative filtering recommender for unexplicit feedback based on the non-linear probabilistic. Technically the authors used variational autoencoders (VAEs) and a generative structure with multinomial probability and Bayesian reasoner for parameter calculation.

Li et al. [68] used different supplemental data such as item information, product tags, and shopping records to solve the data sparsity problem. The authors employed the autoencoder structure for every information source separately to have a better performance.

## 3.2 | Other Deep Learning Architectures in Recommender Systems

To make the study more comprehensive and provide a better understanding of the subject, in the following paragraphs, we present a brief explanation about the other deep learning architectures in recommender systems without limiting the references only to the collaborative filtering approaches.

### 3.2.1 | Recurrent neural networks (RNN)

Recurrent Neural Networks (RNN) are proper solutions for problems related to the changes in data patterns over time. The RNN has a feedback module that makes prediction possible for future input data. From the technical point of view, in a feed-forward neural network, signals flow in only one direction from input to output, one layer at a time. In RNN, the output of a layer is added to the next input and feedbacks to the same layer, which is typically the only layer in the entire network. The sequential pattern and the feature of changes in the hidden layer based on the information opened up RNN to various applications [69]. *Fig. 7,* shows an illustration of the RNN architecture.

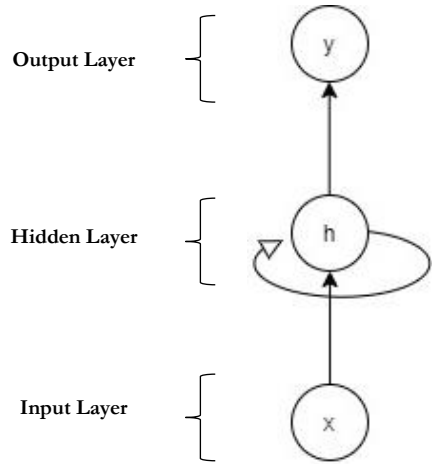

**Fig. 7. A Recurrent Neural Network architecture.**

While there are studies about using RNN in different fields such as recommender systems, however, due to our comprehensive research, compared to the RBM and Autoencoders, fewer studies employed the RNN to build a collaborative filtering system. Ko et al. [70], Proposed a collaborative RNN recommender system that uses a combination of contextual information with latent factors of user preferences to provide more accurate recommendations.

### 3.2.2 | Convolutional neural networks (CNN)

Convolutional Neural Network (CNN) is one of the most dominant deep learning architectures in image processing and machine vision space [71; 72]. From a technical view, CNN is a kind of feed-forward neural network. However, in contrast to the typical neural networks, the convolution operation is used instead of ordinary matrix multiplication in CNN, the system convolves the input data, usually an array from pixels of an image, and analyzes the file, pixel by pixel, to detect edges and extract visual features. The depth of the network and matrixes dimensions vary due to the chosen architecture by the expert [73]. *Fig. 8,* shows a sample of the CNN architecture.

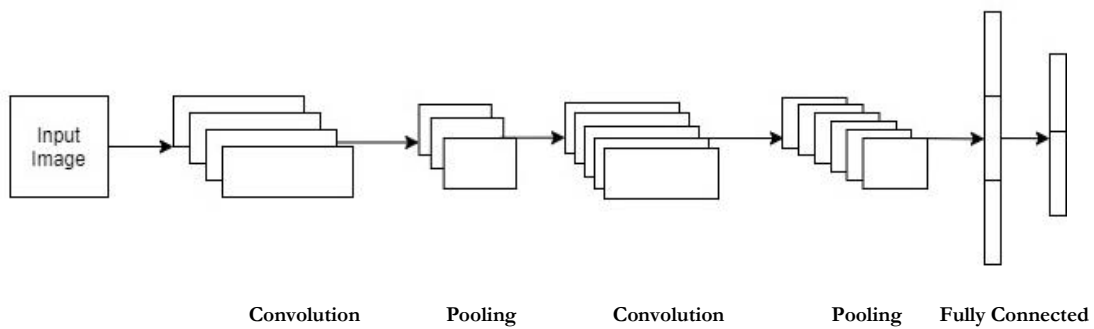

**Fig. 8. A sample of the Convolutional Neural Network architecture.**

Zhang et al. [74], to solve the sparsity problem in collaborative filtering recommenders, introduced Collaborative Knowledge Base Embedding (CKE). The system works based on a hybrid CNN and Autoencoders architectures model to identify images' visual features. The authors also leveraged knowledge-based approaches to consider the dataset's content and textual information to provide a more robust user-item interaction space.

Low et al. [75] proposed a CNN collaborative filtering recommender. The system uses the matrix factorization concept and creates connections between users and items.

Lee et al. [76] proposed a collaborative-based recommender engine that uses audio-visual features to calculate the similarity between videos. The system works based on the CNN architecture to extract visual points and can be highly profitable in video-sharing platforms.

He et al. [77] suggested utilizing a CNN to develop Neural Collaborative Filtering (NCF). The authors named their model ConvNCF. In the proposed structure to present the relation between users and items, the outer product was used alternatively of the dot product so that the system could catch the high-order similarities between defined aspects.

### 3.2.3 | Multilayer perceptron (MLP)

Multilayer Perceptron (MLP) is one of the basic architectures of neural networks. The simple MPL consists of three layers which are input, hidden, and output layers. Each node in a MLP network is known as a perceptron. Obviously, the basic structure of MLP can not be considered as a deep neural network, while employing multiple hidden layers is a technique to develop the architecture and increase its potentials. The scheme of a MLP network is shown in *Fig. 9*. A MLP is an option to convert a linear technique of recommender system into a none-linear system. The MPL-based systems usually are used for supervised models. As a result of being a feed-forward network and having backpropagation, the system continuously modifies the weights and biases to improve the accuracy and achieve the expected result [78].

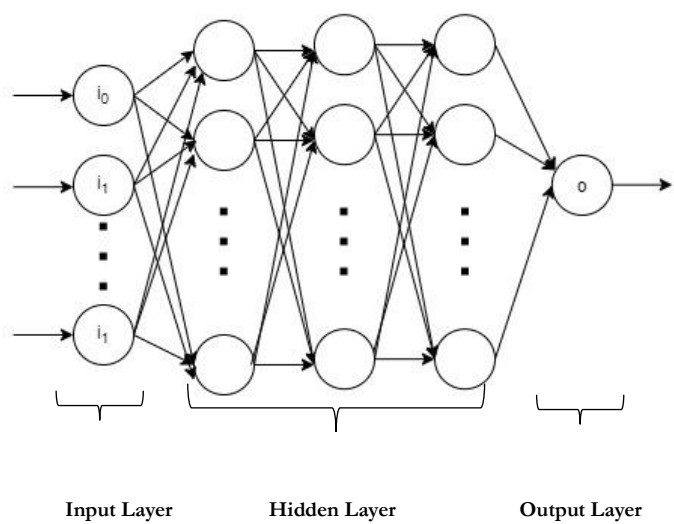

**Fig. 9. A schema of a Multilayer Perceptron architecture with multiple hidden layers.**

Alizadeh [79] proposed a hybrid recommendation system based on the MLP network and collaborative filtering concept. The authors addressed the cold start problem by leveraging the artificial neural network and content-based technique to utilize mutual information from users and items in the dataset. He et al. [80] used MLP as a neural network architecture to build a collaborative filtering recommender system. The authors considered the user-item interaction function, and by leveraging deep learning advantages, improved the basic matrix factorization and collaborative filtering and techniques.

### 3.2.4 | Deep belief networks (DBN)

Combining RBMs together builds a more robust model which can solve the problems efficiently. The model is known as a Deep Belief Network (DBN). In a DBN, the hidden layer of each RBM is the visible layer of the next RBM; this relation continues in the same way for the next RBMs [81]. The last layer in a DBN can be used for clustering or classification. The conceptual architecture of a DBN network is shown in *Fig. 10*.

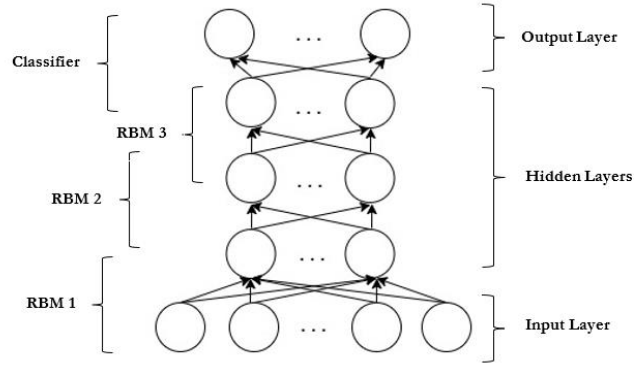

Fig. 10. A basic structure of Deep belief networks.

DBN generally combines both supervised and unsupervised learning steps. Compared with other deep learning architecture, DBNs need less labeled input data; this feature made DBNs a successful solution to implement real-world applications. Moreover, as DBN benefits from a deep network and multiple hidden layers, the system can provide more accurate results, especially compared to shallow nets [82]. Zhao et al. [83], to tackle the sparsity challenge in collaborative filtering recommenders, introduced a hybrid system. The authors employed DBNs to discover user's characteristics. K-nearest neighbor technique is also used to select proper users and execute predictions.

### 3.2.5 | Attentional networks

The attention concept in computer science is formed based on the human ability to concentrate on a specific section of characteristics to perceive the target segments' value. With recent development in deep learning, Attentional Networks have become one of the popular topics in image processing, speech recognition, Natural Language Processing, etc., and also recommender systems [84; 85].

Bahdanau et al. [86] proposed a model machine translation base on the sequence-to-sequence encoder-decoder approach. *Fig. 11,* shows a graphical view of the proposed model

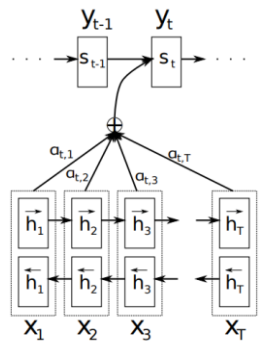

Fig. 11. Graphical view of proposed attention-aware model for translation [86]

In *Fig. 11,* the system's input is a sentence, and the output is its translation. To increase the result's similarity with human translation, the system tries to emphasize more on some specific input parts. $X_1$ to $X_T$ and $h_1$ to $h_T$ inside the rectangles show recurrence step activations. The $a_{t,1}$ to $a_{t,T}$ show how much attention is considered for the related input. Then the sum of the weights provides results boxes on top of the figure. $S_t$ indicates the first time that generated $y_1$, etc.

Tay et al. [87] presented a memory-based Attentional networks architecture for collaborative metric learning. The authors introduced their model as LRML (Latent Relational Metric Learning). In the proposed system, user-item interactions were used as the information resource for the attention module. Jhamb et al. [88] implemented a contextual recommender system based on autoencoder neural network architecture and context-driven attention. The authors used the Attentional network to encode the contextual features into the hidden presentation of the user's characteristics.

### 3.2.6 | Generative adversarial networks (GAN)

A Generative Adversarial Networks (GAN) typically consists of two main neural networks: generator and discriminator. The generator produces new samples of information, and the discriminator validates the generated data for its authenticity. GAN architecture became so popular in image processing deep learning-based systems [73; 89]. Tong et al. [90] proposed a Collaborative Generative Adversarial Network (CGAN) to build a recommender system. The authors used an auto-encoder as a generator module that takes features from user activities about items. Moreover, adversarial training was employed to enhance system efficiency and productivity.

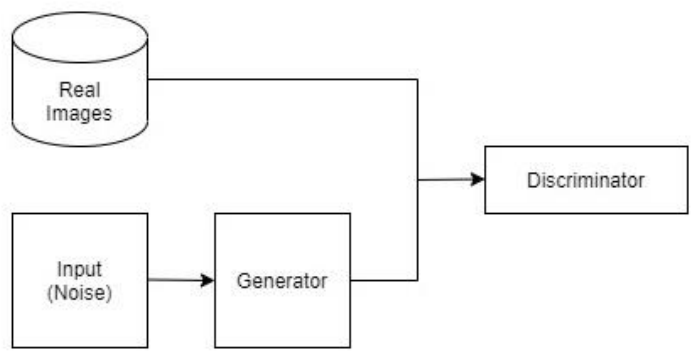

**Fig. 12. A fundamental structure of the Generative adversarial network.**

*Fig. 12,* shows a GAN which produces fake images. The network has two inputs: a dataset of real images (top input) and a D-dimensional noise vector (bottom input). Generator Component produces fake images. In the next step, samples from real images and fake images become validate by the discriminator.

## 4 | Discussion

In this section, we discuss our work results and analyze the details from some of the most important views that can be applied to the topic. From *Table 4,* it is observed that Autoencoders and Restricted Boltzmann Machine (RBM) are the most popular deep learning architecture for building collaborative filtering-based recommender systems; and other approaches such as Convolutional Neural Network (CNN), Recurrent Neural Network (RNN), Deep belief networks (DBN) Multilayer Perceptron (MLP), Attentional networks and Generative adversarial networks (GAN) have not received enough attention through the reviewed studies. Although, there could be different reasons for the popularity of Autoencoders and RBM, based on the provided data in *Table 5*, one of the critical differences is their training type which is Unsupervised for both. Moreover, their potentials in common applications such as feature extraction and reducing dimensions are highly compatible with collaborative filtering recommender systems structure.

**Table 4. Literature on using deep learning architectures in collaborative filtering recommender systems.**

| Deep learning Architecture | Publication | Dataset |
|---|---|---|
| Restricted Boltzmann Machines (RBM) | Louppe. [48] | Netflix |
| | Georgiev and Nakov. [49] | MovieLens |
| | Liu et al. [50] | MovieLens |
| | Zheng et al. [51] | MovieLens, Netflix |
| | Jia et al. [52] | Custom, Renren, Meetup |
| | Du et al. [53] | MovieLens |
| | Wu et al. [54] | MovieLens |
| Autoencoders | Ouyang et al. [56] | MovieLens |
| | Li et al. [57] | MovieLens, Book-Crossing, Advertising |
| | Sedhain et al. [58] | MovieLens, Netflix |
| | Strub and Mar. [59] | MovieLens, Jester |
| | Wang et al. [60] | Netflix, CiteULike |
| | Wang et al. [61] | Netflix, CiteULike |
| | Ying et al. [62] | CiteULike |
| | Wei et al. [63] [64] | Netflix |
| | Suzuki and Ozaki. [65] | MovieLens |
| | Li and She. [66] | CiteULike |
| | Liang et al. [67] | MovieLens, Netflix, Million Song |
| | Li et al. [68] | MovieLens, OfflinePay |
| Recurrent Neural Networks (RNN) | Ko et al. [70] | Brightkite, LastFM |
| Convolutional Neural Networks (CNN) | Zhang et al. [74] | MovieLens, IntentBooks |
| | Lo et al. [75] | MovieLens, Pinterest |
| | Lee et al. [76] | MovieLens, YouTube |
| | He et al. [77] | Yelp, Gowalla |
| Multilayer Perceptron (MLP) | Divan and Alizadeh. [79] | MovieLens, Netflix |
| | He et al. [80] | MovieLens, Pinterest |
| Deep Belief Networks (DBN) | Zhao et al. [83] | MovieLens |
| Attentional networks | Tay et al. [87] | MovieLens, Netflix, IMDb, LastFM, Books, Delicious, Meetup, Twitter |
| | Jhamb et al. [88] | MovieLens, Meetup |
| Generative Adversarial Networks (GAN) | Tong et al. [90] | MovieLens, Netflix |

*Table 5* provides a condensed review and comparison of the different deep learning architectures. It has to be mention that some values, such as examples of the Common Applications presented in the table, also could be implemented in hybrid applications. Likewise, while RBMs are considered as generative models and "Unsupervised" chose as their "Training Type", they can have components of the discriminative model and do their training phase in a supervised system [25].

**Table 5.** Deep learning architectures comparison.

| Deep learning Architecture | Training Type | Training Algorithm | Common Applications |
|---|---|---|---|
| **Restricted Boltzmann Machines (RBM)** | Unsupervised | Gradient Descent | Feature Extraction |
| **Autoencoders** | Unsupervised | Backpropagation | Encoding; Reducing Dimensions |
| **Recurrent Neural Networks (RNN)** | Supervised | Gradient Descent / Backpropagation | Natural Language Processing; Translating Languages |
| **Convolutional Neural Networks (CNN)** | Supervised | Gradient Descent / Backpropagation | Image Processing |
| **Multilayer Perceptron (MLP)** | Supervised | Gradient Descent / Backpropagation | Stochastic Solution; Fitness Approximation |
| **Deep Belief Networks (DBN)** | Supervised | Gradient Descent | Classification; Anomaly Detection |
| **Attentional Networks** | Supervised | Gradient Descent / Backpropagation | Image Processing; Speech Recognition, Natural Language Processing |
| **Generative Adversarial Networks (GAN)** | Unsupervised | Backpropagation | Generating Data; Reconstruction Data and Images |

We also classified papers based on their datasets. *Table 6*, shows datasets that researchers frequently used for building deep learning collaborative filtering recommender systems. The values of this table demonstrate that MovieLens [91] is the most common dataset in building collaborative filtering-based recommender systems that used deep learning architectures. MovieLens is a result of the GroupLens research project that the University of Minnesota did. It is a movie recommender website that users can rate the movies from 1, which means the worst score to 5, which means the maximum satisfaction. Two versions of the dataset were used in the reviewed studies, entitled MovieLens 100k and MovieLens 1M. Respectively, MovieLens 100k includes 100,000 ratings for 1,682 movies from 943 users, and MovieLens 1M includes 6,040 users who rated 1,000,000 ratings for 3,952 movies.

Table 6. Frequency of used datasets in the reviewed papers.

| Title | Frequency |
|---|---|
| MovieLens | 22 |
| Netflix | 10 |
| CiteULike | 4 |
| Meetup | 3 |
| Pinterest | 2 |
| LastFM | 2 |
| YouTube | 1 |
| Twitter | 1 |
| IMDb | 1 |
| Jester | 1 |
| Book-Crossing | 1 |
| Brightkite | 1 |
| Books | 1 |
| IntentBooks | 1 |
| GowaYelplla | 1 |
| Delicious | 1 |
| OfflinePay | 1 |
| Million Song | 1 |
| Renren | 1 |
| Advertising | 1 |
| Customized datasets | 1 |

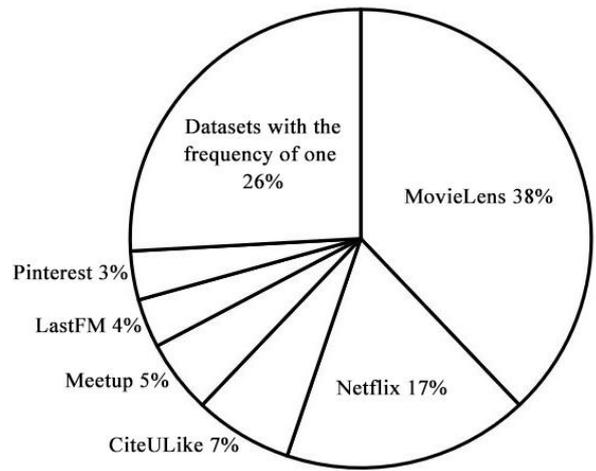

Fig. 13. Diversity of used datasets in the reviewed papers.

*Fig. 13,* shows a pie chart that presents another view of the given values in *Table 6*. However, to make a better presentation, datasets with the frequency of one, aggregated in one group, so the chart is divided into seven parts. Again, it is clear that the MovieLens dataset is the most common dataset in building deep learning collaborative filtering recommenders. Another categorization in this work was done based on the publishing date of the reviewed papers; *Fig. 14,* shows the diversity of these studies based on their publishing date.

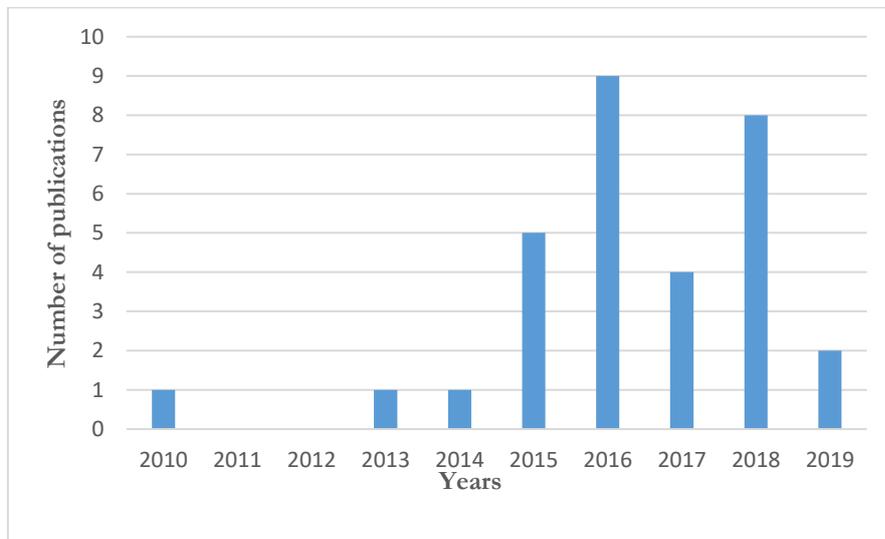

Fig. 14. Diversity of reviewed papers based on their publishing date.

As shown in *Fig. 14*, the year 2016 has the most significant number of published papers about collaborative filtering-based recommender systems built based on deep leering architectures. Respectively, 2018 and 2015 are the second and third ones. As a consequence of these values, it can be concluded that there is a direct relationship between the studies about recommender systems and achievements in providing efficient deep learning architecture in recent years.

### 4.1 | Why Deep Learning Architectures for Recommendation?

With the tremendous extension in the volume of online data, handling the user's requests with traditional information retrieval and decision support systems has become highly challenging. However, as deep learning techniques are efficient in combining multiple sources of information and explore hidden features and patterns from them, big data caused increasing the popularity of these systems in recent years [16].

Another noticeable strength of employing deep neural networks, especially in collaborative filtering-based recommender systems, is the capability in transforming the multi-dimensional user-item matrixes with sparsity into a smaller matrix with more data. For instance, Unger et al. [92] employed autoencoders to decrease the dimensions of environmental features and overcome the sparsity in context-aware recommendation systems.

Extracting visual features by convolutional neural networks as complemental data to user's history and ratings is an adequate technique to solve the cold start problem [93]. Shin et al. [94] proposed a blog recommender system. The authors to deal with the cold-start problem combined derived features from textual data and pictures by CNN.

As mentioned in section 2.2.3, identifying grey sheep users is another challenge ahead of building profitable recommender systems; due to the advantages of neural networks for clustering data based on different features. For instance, Rabba [95] proposed a system that detects grey sheep users based on unsupervised learning clustering techniques.

## 5 | Conclusion

Nowadays, practical filtering information and providing personalized recommendations have become increasingly critical, notably in online-based industries such as e-commerce or customer services. By increasing the interest in applying deep learning in different fields, enhancing recommender systems' performance by utilizing these kinds of approaches has also become increasingly pervasive.

In this study, we provided an extensive review of utilizing and leveraging deep learning architectures in collaborative filtering recommender systems. However, in contrast to the subject's prior works that reviewed different techniques generally, we specifically provided a comprehensive review of deep learning-based collaborative filtering recommender systems. We chose collaborative filtering as the most common technique in building recommender systems and tried to clarify its relationship with deep learning architectures. Moreover, another analysis was done based on the researchs' datasets. Due to the results, the MovieLens dataset is the most popular dataset to make a collaborative filtering recommender system that was build based on deep learning architectures.

Due to the results, Autoencoders and Restricted Boltzmann Machine (RBM) are the most popular ones, and other architectures such as Recurrent Neural Networks (RNN), Convolutional Neural Networks (CNN), Multilayer Perceptron (MLP), Deep Belief Networks (DBN), Attentional networks and Generative Adversarial Networks (GAN), gained fewer attention from researchers. In future work, we are interested in study the results of applying deep learning in varied fields of recommenders and compare the degree of influence from different perspectives, such as revenue, engagement, etc.